\documentclass[letterpaper]{article} 
\usepackage{aaai24}  
\usepackage{times}  
\usepackage{helvet}  
\usepackage{courier}  
\usepackage[hyphens]{url}  
\usepackage{graphicx} 
\usepackage{natbib}  
\usepackage{caption} 
\usepackage{algorithm}
\usepackage{algorithmic}
\usepackage{amsmath}

\usepackage{newfloat}
\usepackage{listings}
\DeclareCaptionStyle{ruled}{labelfont=normalfont,labelsep=colon,strut=off} 
\floatstyle{ruled}
\newfloat{listing}{tb}{lst}{}
\floatname{listing}{Listing}
\title{Inverse Design of Photonic Crystal Surface Emitting Lasers is a Sequence Modeling Problem}

\author {
    Ceyao Zhang\textsuperscript{\rm 1,\rm 2}\equalcontrib,
    Renjie Li\textsuperscript{\rm 2,\rm 3}\equalcontrib,
    Cheng Zhang\textsuperscript{\rm 2}\equalcontrib,
    Zhaoyu Zhang\textsuperscript{\rm 2,\rm 3}\footnotemark[2],
    Feng Yin\textsuperscript{\rm 2}\thanks{Corresponding author}
}
\affiliations {
    \textsuperscript{\rm 1}Future Network of Intelligence Institute (FNii), The Chinese University of Hong Kong, Shenzhen\\
    \textsuperscript{\rm 2}School of Science and Engineering (SSE), The Chinese University of Hong Kong, Shenzhen\\
    \textsuperscript{\rm 3}Shenzhen Key Laboratory of Semiconductor Lasers\\
    \{ceyaozhang2@link., renjieli@link., zhangzy, yinfeng\}@cuhk.edu.cn
}

\begin{document}
\maketitle
\begin{abstract}
Photonic Crystal Surface Emitting Lasers (PCSEL)'s inverse design demands expert knowledge in physics, materials science, and quantum mechanics which is prohibitively labor-intensive. Advanced AI technologies, especially reinforcement learning (RL), have emerged as a powerful tool to augment and accelerate this inverse design process. By modeling the inverse design of PCSEL as a sequential decision-making problem, RL approaches can construct a satisfactory PCSEL structure from scratch. However, the data inefficiency resulting from online interactions with precise and expensive simulation environments impedes the broader applicability of RL approaches. Recently, sequential models, especially the Transformer architecture, have exhibited compelling performance in sequential decision-making problems due to their simplicity and scalability to large language models. In this paper, we introduce a novel framework named \textit{PCSEL Inverse Design Transformer} (PiT) that abstracts the inverse design of PCSEL as a sequence modeling problem. The central part of our PiT is a Transformer-based structure that leverages the past trajectories and current states to predict the current actions. Compared with the traditional RL approaches, PiT can output the optimal actions and achieve target PCSEL designs by leveraging offline data and conditioning on the desired return. Results demonstrate that PiT achieves superior performance and data efficiency compared to baselines.
\end{abstract}

\section{Introduction}


Photonic Crystal Surface Emitting Lasers~\citep[PCSELs;][]{hirose2014watt, yoshida2019double, noda2017photonic} are a type of nanoscale laser that combines the benefits of photonic crystals~\citep[PhC;][]{quan2010photonic} and Vertical Cavity Surface Emitting Lasers~\citep[VCSELs;][]{chang2000tunable}. PhCs are artificial structures that have a periodic refractive index modulation in semiconductor materials. This periodicity creates a photonic bandgap that inhibits the propagation of light in certain frequency ranges to amplify the optical resonance effect. VCSELs are lasers that emit light perpendicular to the surface of the semiconductor structure, which allows for efficient coupling to optical fibers and other optical components. PCSELs combine these two technologies to create advanced lasers that have several advantages over traditional ones and therefore enjoy the best of both worlds. PCSELs have great potential for important applications in sensing, autonomous driving, medicine, machining, and telecommunication. Fundamentally, when an electrical pumping current is injected into the active layer (that is, the core layer of PCSEL), it emits laser light which is then confined and amplified within the PhC resonant cavity~\citep{sze2021physics}. Additionally, the active layer may contain quantum wells that increase the recombination rate of spontaneous photon emission (which was first predicted by Albert Einstein in his quantum mechanics papers~\citep{hilborn1982einstein}) and thus substantially enhance the lasing effect~\citep{sze2021physics}. So the bottom line is that the PhC layer is used to control the amplitude and direction of the emitted light, while the active layer is what actually generates the light. Therefore, proper design of the PhC layer and the active layer plays a central role in the overall quality of a PCSEL, which brings to us the critical yet challenging PCSEL inverse design problem. This problem is a combination of disciplines including physics, materials science, and quantum mechanics which generally demands prohibitively high-level domain knowledge.

\begin{figure}[htbp]
  \centering
  \includegraphics[width=\linewidth]{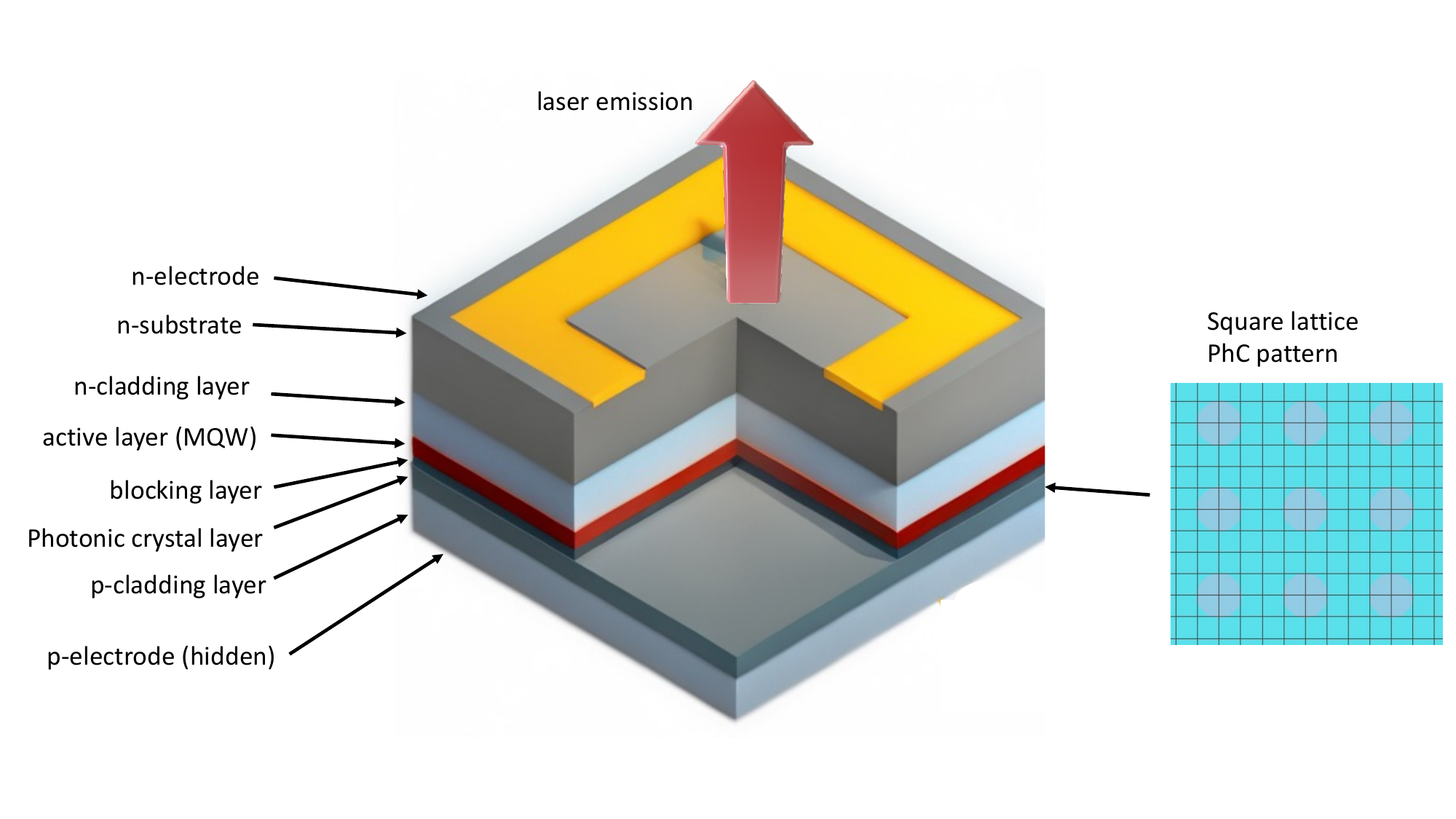}
  \caption{PCSEL structure schematic, with the active layer and PhC cavity layer shown. Based on the band-edge mode, a large area lasing resonance mode is formed within the PhC resonance cavity. Lasing arises from the evanescent coupled MQW gain medium in the active layer. The inverse design problem therefore focuses on optimizing the designs of the active layer and the PhC cavity as a whole. The size of the device is about 200 microns in side lengths and 400 microns in height and solely manufactured with semiconductor materials.}
\end{figure}

Recent developments in Transformer~\citep{vaswani2017attention} architectures witness the blooming performance from prediction tasks into decision-making problems~\citep{wen2023large}. 
This advancement, namely, sequential decision-making as sequence modeling, is notably distinct from the traditional approaches in reinforcement learning (RL), which are typically focused on learning a singular policy for a specific, narrowly defined behavior distribution.
Given the broad range of successful implementations of these models, this study aims to explore their applicability to the inverse design of PCSEL.

In this paper, we introduce a novel framework named PCSEL Inverse Design Transformer (PiT) that abstracts the inverse design of PCSEL as a sequence modeling problem. The central part of our PiT is a Transformer-based structure that leverages the past trajectories and current states to predict the current actions. Compared with the traditional RL approaches, PiT can output the optimal actions and achieve target PCSEL designs by leveraging offline data and conditioning on the desired return. Results demonstrate that PiT achieves superior performance and data efficiency compared to baselines.





\section{Related Works}

\paragraph{Optimzing the PCSEL through AI} 
Overall, recent advancements in machine learning~\citep{goodfellow2016deep} and optimization algorithms~\citep{wiecha2017evolutionary} have propelled the progress of inverse designs in science. Early in the 90s, heuristic, evolutionary~\citep{hegde2019photonics}, and gradient-based~\citep{zhang2020single} optimization algorithms began to emerge prolifically. Key algorithms include simulated annealing~\citep{bertsimas1993simulated}, Newton's method~\citep{milzarek2014semismooth}, Bayesian optimization~\citep{shahriari2015taking}, Monte Carlo method~\citep{rubinstein2016simulation}, particle swarm~\citep{ma2020parameter} and genetic algorithm~\citep{ren2021genetic} etc. 
These algorithms provide a new way of thinking when facing hard non-convex optimization problems, which act as a solid foundation for scientific problems. 
However, the issue remains of heavy human involvement due to sophisticated trial-and-error iterations. To solve this, in around 2012, researchers proposed deep learning (DL)~\citep{krizhevsky2012imagenet,goodfellow2016deep} frameworks to construct a mapping relationship between input data and output targets through Deep Neural Networks. 
In particular, DL consists of supervised, unsupervised, and reinforcement learning (RL)~\citep{sutton2018reinforcement}. These DL models greatly bolstered the efficiency of inverse design in science, pushing the possibility of automated design into a new era~\citep{so2020deep, jiang2021deep, li2021deep, mirhoseini2021graph, li2022smart, degrave2022magnetic, li:hal-04175312, kuprikov2022deep}. 
Circa 2023, a novel framework based on RL, called Learning to Design Optical-Resonators (L2DO)~\citep{li2023deep}, provides the solution for autonomous inverse design of nanophotonic chips without human intervention. With two orders of magnitude higher sample efficiency compared to supervised learning, L2DO has preliminarily realized RL-driven chip inverse design on an algorithmic level. However, to the best of our knowledge, there aren't any published results on the inverse design of PCSELs via AI methods to this day. Due to the strategic significance of PCSELs for a host of key industries, we believe there is an urgent need to develop an RL-based approach for PCSEL's rapid inverse design.

\paragraph{Sequential Models for Sequential Decision-Making}
Recent research on transformer models for sequential decision-making has revolutionized the field by pushing boundaries and introducing novel capabilities. 
Transformers~\citep{wen2023large} excel in capturing long-range dependencies, efficiently assigning credit, and modeling temporal dynamics. 
These have significantly advanced the field and hold promise for applications in reinforcement learning. Foundational works in the application of transformers to sequential decision-making include Decision Transformer (DT)~\citep{chen2021decision} and Trajectory Transformer(TT)~\citep{janner2021offline}. 
These works have made a significant impact on the field by introducing transformative methodologies.
DT, for instance, revolutionized traditional offline RL approaches by employing a transformer decoder as the backbone model. It uses (return-to-go, state) as input data, and chooses action as output. This method outperforms the offline RL baseline CQL. 
TT is another foundational application of transformers in sequential decision-making, which utilizes the GPT model~\citep{radford2018gpt} as its backbone, demonstrating remarkable performance in long-range planning tasks.
Other efforts like UPDeT~\citep{hu2021updet}  for multi-agent planning and MGDT~\citep{lee2022multi} for multigame decision-making highlight the versatility of transformers. 
Looking ahead, transformer models show great promise to transform decision-making in the coming years by enabling a more nuanced understanding of situational context and more predictive recommendations. 

\section{Background}
\subsection{Modeling Inverse Design of PCSEL as a Sequential Decision-making problem}

Sequential decision-making describes a situation where the decision-maker makes successive observations of a process before a final decision is made. In most sequential decision problems, there is an implicit or explicit cost/regret/reward associated with each observation or action. The procedure to decide when to stop taking observations and when to continue is called the ‘stopping criteria’. The objective of sequential decision-making is to find a stopping criterion that optimizes the decision in terms of minimizing losses or maximizing returns, including observation costs. The optimal stopping criteria are also called optimal strategy and optimal policy, which are commonly adopted in classic RL algorithms. 

    

To the best of our knowledge, \citet{li2023deep} is the first work on modeling the inverse design of PCSELs as a Sequential decision-making problem, which offers a structured and efficient approach to navigating the complex design space. 
Sequential Decision-Making involves breaking down the design process into a series of decisions, each contingent upon the outcomes of preceding ones. This approach is particularly suited to PCSEL design due to the layered nature of their construction and the interdependence of various design parameters. 
There are a few key components of the sequential decision-making model.
\begin{itemize}
    \item State Space represents the current status of the PCSEL design, encompassing all relevant parameters (e.g., layer thicknesses, refractive indices, geometric patterns).
    \item Action space is discrete and consists of 16 actions. Each action represents an increase or decrease in geometric parameters by one unit according to the scale of the parameters. For example, action 0 is to increase parameter 0 by 25, and action 3 is to decrease parameter 1 by 2.5. 
    \item Reward: As to the quality of a PCSEL, there are a few indicators to determine its performance, which we listed in Table~\ref{table:indicators}. 
    Unify the dimensions between different indicators through weighting, we can calculate the score from the parameters returned at each step, as in Eqn~\ref{equ:score}. 
    In PCSEL, the reward function is the difference between the score obtained in the current step and the score obtained in the previous step. 
    As the first step, we take the score as a reward. To analyze the composition of the score in more detail, we can examine the parameters returned by the environment and the method of calculation.
\end{itemize}
\begin{table}[h]
    \centering
    \begin{tabular}{|c|c|}
  \hline
  notations & Indicators \\
  \hline
   $Q$ & Q-factor \\
   \hline
   $lam$ & lambda\\
   \hline
   $power$ & power\\
   \hline
   $area$ & area\\
   \hline
   $div\_angle$ & divergence angle \\
   \hline
    \end{tabular}
    \caption{
    Indicators that determine PCSEL performance.
    }
    \label{table:indicators}
\end{table}

\begin{gather}
    r_1 = 1 - (Q_{goal} - Q) / Q_{goal} \\
    r_2 = 1 - |lam_{goal} - lam| / lam_{goal} \\
    r_3 =1 - (area_{goal} - area) / area_{goal} \\
    r_4 = 1 - (power_{goal}- power) / power_{goal} \\
    r_5 = 1 + (div\_angle_{goal} - div\_angle) / div\_angle_{goal} \\
    score=\gamma * r_1+\epsilon *r_2+\beta * r_3+\alpha *r_4+\eta *r_5  \label{equ:score}
\end{gather}

The designer continues to take actions to set/update the layout until a pre-set stopping criteria is met, 
which in our case is the target PCSEL performance characteristics listed in Table 1. 
When the target characteristics are met, one can consider the cumulative return is maximized or an optimal policy is found. Due to the high-level domain knowledge and labor intensity required by PCSEL inverse design, the authors believe modeling it as a sequential decision problem can largely alleviate human labor and accelerate the R\&D of advanced PCSEL lasers.


\section{Proposed Methods}

\subsection{Inverse Design of PCSEL is a Sequential Modeling Problem}





\begin{figure}[htbp]
  \centering
  \includegraphics[width=0.6\linewidth]{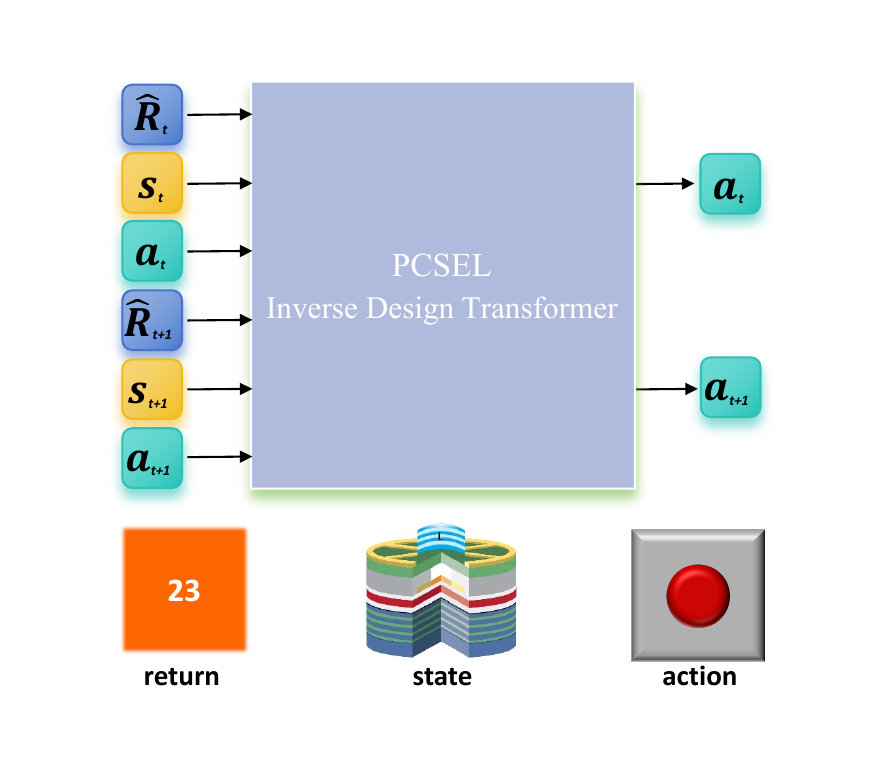}
  \caption{The architecture of PiT.}
\end{figure}

The PCSEL inverse design can be conceptualized as a sequential decision model, where the goal is to maximize the score within a limited number of steps through strategic adjustments. This model is characterized by a clear reward target, a small and discrete action space, and the Markov property. Given these properties, it is naturally suited for RL algorithms.

However, unlike games that can be quickly simulated, data collection for PCSEL inverse design is challenging due to its complexity. Therefore, considering data efficiency, offline RL is a more suitable choice. 
Our PiT model is based on sequence models, for example, DT~\citep{chen2021decision} or TT~\citep{janner2021offline}. 
Among the existing sequence models for sequential decision-making, we choose DT for its simplicity and superior performance.



\subsection{Dataset}

\begin{figure}[htbp]
  \centering
  \includegraphics[width=0.95\linewidth]{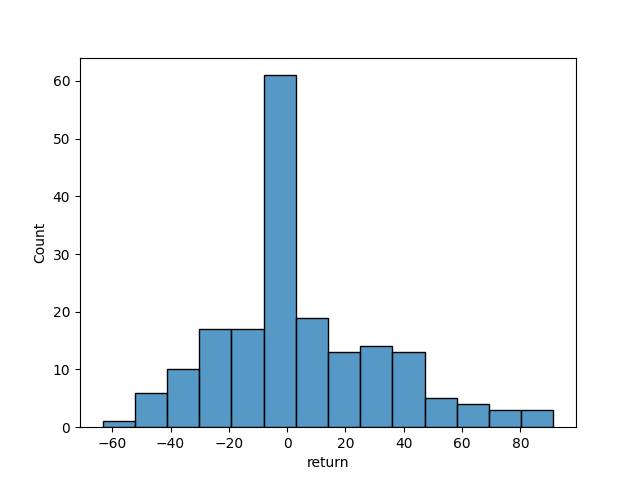}
  \caption{The frequency of the total return of the trajectories used to train PiT.}
\end{figure}

The dataset of offline trajectories for training our PiT is extracted from replay buffers that were saved during previous online RL experiments associated with PCSELs. The offline dataset contains roughly 16,057 samples (each sample includes state, action, reward, and next state). 
Specifically, the state is defined as a vector of design parameters of the PCSEL, the action is a change/update in the state, and the reward is defined as how close we are to the target PCSEL design. For instance, if we wish to achieve a target PCSEL design with a Q-factor of 5 million, a wavelength (lambda) of 1310 nm, a power conversion rate of 80\%, a lasing area of 3.0e-13 square meters, and a divergence angle of 1.0 degrees, we would train PiT to inverse design a PCSEL that meets these target characteristics. 
The actual criteria is a weighted sum of these five reward of merit according to Eqn. 1-6 and is referred to as a score in the remaining text.


\section{Experiments}

In this section, we performed a comprehensive performance comparison involving our PiT and behavior cloning.
Models were trained on both the whole dataset and a subset comprising exclusively of the rising dataset, where the score at the last timestep exceeded the initial timestep.
\subsection{Baseline}
We choose behavior cloning as our baseline. Behavior cloning involves training a network to imitate the behavior of a demonstrated or an expert dataset. 
The network is an MLP with 3 layers and 256 embedding dimensions. 
It is a form of supervised learning where the model learns to map observations to actions directly by mimicking the expert's actions.




\begin{figure}[tbh]
  \centering
    \centering
    \includegraphics[width=\linewidth]{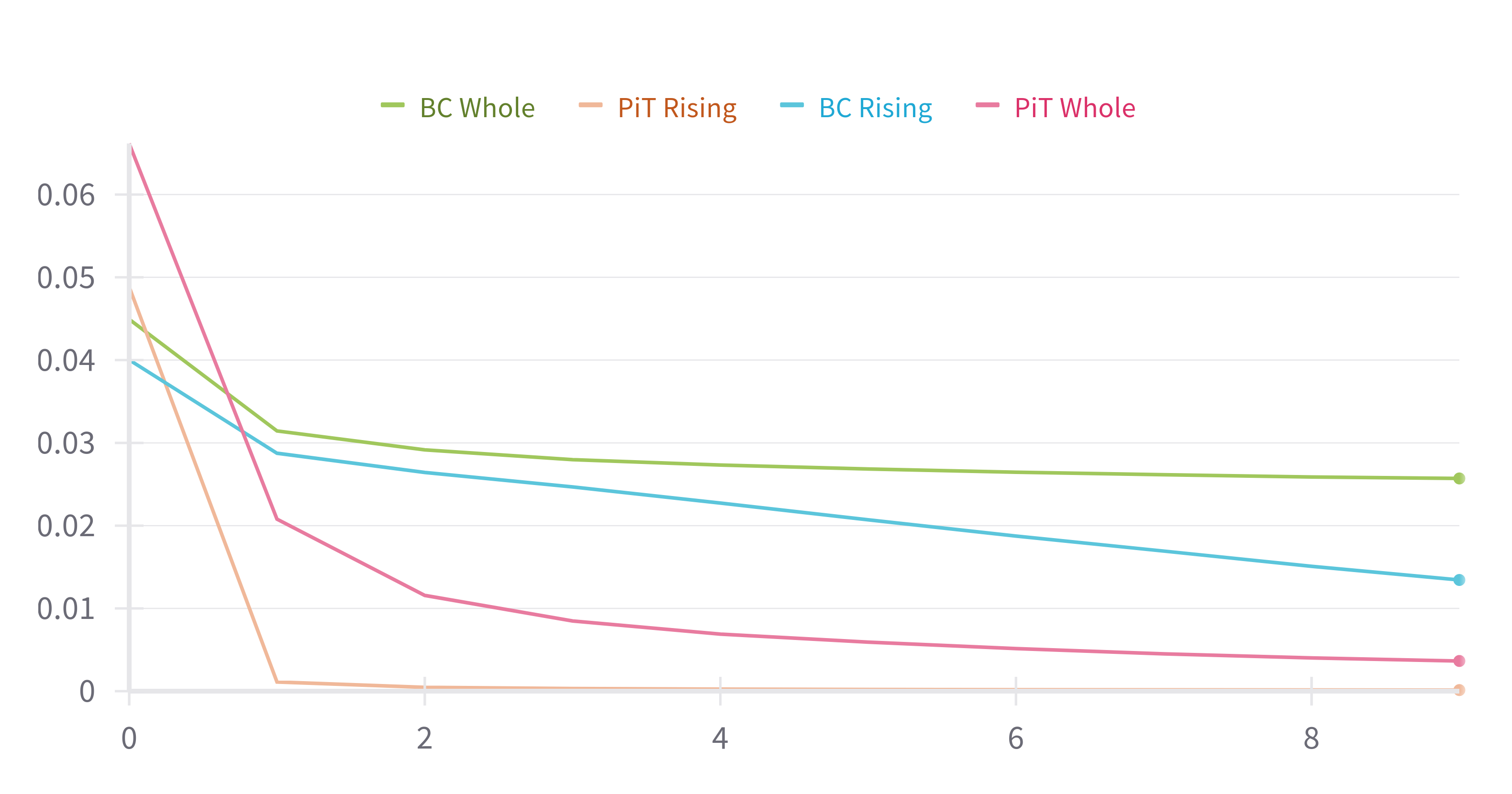}
    \caption{Training loss curve of both BC and PiT under the whole and rising dataset.}
    \label{fig:training_loss}
\end{figure}

From Figure~\ref{fig:training_loss}, it is evident that the PiT exhibits a lower training loss compared to baseline BC. 
This lower training loss indicates that PiT has a greater ability to predict the next action accurately on the training set.


\subsection{Results}

In Table~\ref{table:all_methods}, we compare PiT's results to the best literature baselines. The results are quantified by a score which is defined in Eqn. 6. 
Moreover, thanks to the offline RL framework, PiT obtained better data efficiency than the baselines as it doesn't require online environment interactions or labor-intensive manual optimizations by human designers. Therefore, we conclude that PiT has acquired state-of-the-art capabilities for PCSEL inverse design with consistently better performance than baselines. Nonetheless, we recognize that PiT's score is still quite some distance away from the target, which means there's room for improvement in future research.

\begin{table}[htbp]
\centering
\begin{tabular}{|c|c|}
\hline
\citet{imada1999coherent}  &  9.19709 \\ \hline
\citet{ohnishi2004room}   &  19.2290   \\ \hline
\citet{sakai2005lasing} & 21.23469  \\ \hline
\citet{hirose2014watt}& 18.95918 \\ \hline
\citet{hsu2017electrically} & 10.01160 \\ \hline
\citet{chen2021improvement}& 2.117346 \\ \hline
\citet{wang2021photonic} & 41.8554  \\\hline
\citet{itoh2022high} & 20.2335 \\ \hline
\textbf{BC}  & 71.28  \\\hline
\textbf{PiT} & 73.95 \\ \hline 
\end{tabular}
\caption{PiT's results compared to baselines in the literature.}
\label{table:all_methods}
\end{table}

\subsection{Discussion}
In this section, we conducted a performance comparison between PiT and BC trained on both the whole dataset and the rising dataset. The results are listed in Table~\ref{table:BC-PiT}. 
According to the table, we can see that regardless of whether BC or PiT methods are used, the results based on the rising dataset are better than the results of the whole dataset.
This phenomenon indicates that the offline dataset we choose has a large impact on performance. 
More specifically, the better the data set we use (i.e. the larger the reward), the better the final performance will be.
Therefore, if we can train a better RL policy and collect a dataset with higher returns, it is possible to further improve the performance of PCSEL by a large margin.
\begin{table}[htbp]
  \centering
  \label{tab:performance}
  \begin{tabular}{|c|c|c|}
    \hline
    \textbf{Dataset \textbackslash Method} & \textbf{BC} & \textbf{PiT} \\
    \hline
    Whole  & 68.51  &  70.54 \\
    \hline
    Rising & 71.28 & 73.95 \\
    \hline
  \end{tabular}
  \caption{Performance Comparison for PiT and BC under the whole and rising dataset.}
  \label{table:BC-PiT}
\end{table}

Finally, we discuss the potential limitations of our method as we as future work. First, the effectiveness of the PiT model is heavily dependent on the availability and quality of offline data, which could be a limiting factor in some scenarios where data is scarce.
Second, while effective in the specific context of PCSEL inverse design, it's unclear how well this approach generalizes to other types of photonic devices or design objectives. In the future, we will expand the realm of PiT by demonstrating its effectiveness and versatility for different types of devices. We will also generate a diverse pool of data for photonic inverse design problems and make it open-source.

\section{Conclusion}
In this paper, we investigate the PCSEL inverse design problem through sequential modeling that facilitates the use of offline data and eliminates the need for online interactions. 
Our simulation experiments show the effectiveness of our proposed framework.
We believe that our work points out promising future avenues to design advanced PCSEL lasers and photonics in general.
We further analyze the impact of Transformer structure selection and highlight several potential solutions to improve the performance in the future.

\section*{Acknowledgement}
The work was supported in part by the National Natural Science Foundation of China (NSFC) under Grant No. 62271433, by the Basic Research Project No. HZQB-KCZYZ-2021067 of Hetao Shenzhen-HK S\&T Cooperation Zone, by the National Key R\&D Program of China with grant No. 2018YFB1800800, by NSFC with Grant No. 62293482, by the Shenzhen Outstanding Talents Training Fund 202002, by the Guangdong Research Projects No. 2017ZT07X152 and No. 2019CX01X104, by the Guangdong Provincial Key Laboratory of Future Networks of Intelligence (Grant No. 2022B1212010001), by the Shenzhen Key Laboratory of Big Data and Artificial Intelligence (Grant No. ZDSYS201707251409055), and by Shenzhen Science and Technology Program under Grant No. JCYJ20220530143806016 and No. RCJC20210609104448114, by NSFC under Grant No.62174144, Shenzhen Science and Technology Program under Grant No.JCYJ20210324115605016, No.JCYJ20210324120204011, No.JSGG20210802153540017, and No.JCYJ20220818102214030, Guangdong Key Laboratory of Optoelectronic Materials and Chips under Grant No.2022KSYS014, Shenzhen Key Laboratory Project under Grant No.ZDSYS201603311644527; Longgang Key Laboratory Project under Grant No.ZSYS2017003 and No.LGKCZSYS2018000015; Shenzhen Research Institute of Big Data; President’s Fund (PF01000154). 

\bibliography{aaai24}


\end{document}